



\magnification = 1200
\overfullrule = 0 pt
\baselineskip = 16 pt
\hsize = 5.7 truein
\vsize = 8.75 truein

\font\titolo = cmbx10 scaled \magstep2
\font\abstract =cmr9

\font\autori = cmsl10 scaled \magstep2

\def\CcC{{\hbox{\tenrm C\kern-.45em{\vrule height.67em width0.08em depth-.04em
\hskip.45em }}}}
\def\RrR{{\hbox{\tenrm I\kern-.17em{R}}}}
\def\HhH{{\hbox{\tenrm {I\kern-.18em{H}}\kern-.18em{I}}}}
\def\DdD{{\hbox{\tenrm {I\kern-.18em{D}}\kern-.36em {\vrule height.62em
width0.08em depth-.04em\hskip.36em}}}}
\def\ZzZ{{\hbox{\tenrm Z\kern-.31em{Z}}}}
\def\IiI{{\hbox{\tenrm I\kern-.19em{I}}}}
\def\NnN{{\hbox{\tenrm {I\kern-.18em{N}}\kern-.18em{I}}}}

\def\mapbelow#1{\smash{\mathop{\longrightarrow}\limits_{#1}}}

\hrule height 0 pt
\medskip

\vskip .85 truein
\hrule height 0 pt
\vskip 1.25 truein
\nopagenumbers

\centerline{\titolo Dissipation and memory capacity}
\centerline{\titolo in the quantum brain model}
\vskip .15 truein

\centerline{
{\autori  Giuseppe Vitiello}}
\vskip .15 truein

\centerline{{\abstract Dipartimento di Fisica, Universit\'a di Salerno,
84100 Salerno, Italy}}

\vskip .6 truein

\centerline {\bf ABSTRACT}
\vskip .05 truein
{\abstract   The quantum model of the brain proposed
 by Ricciardi and Umezawa is  extended to dissipative
  dynamics in order to study the problem of memory capacity.
It is shown  that infinitely many
vacua
are accessible to memory printing in a way
that in sequential information recording the storage of a new information
  does not destroy the previously stored ones, thus allowing a huge memory
capacity. The mechanism of information printing is shown to
induce breakdown  of  time-reversal  symmetry.
Thermal
properties of the memory states as well as their relation with
squeezed coherent states are finally discussed.}

\vfill
\line{e-mail ~ vitiello@sa.infn.it ~~~~~Int. J. Mod. Phys.B, in print\hfill}
\eject
\pageno = 2
\footline{\hss \tenrm -- \folio $\,$ -- \hss}
\hrule height 0 pt
\bigskip

\line{{\bf 1. Introduction}\hfill}
\medskip
\indent The purpose of this paper is the study of the problem of memory
capacity in the Ricciardi-Umezawa quantum model of brain[1]
by resorting to
recent results on dissipative systems in quantum field theory (QFT)[2].

Coupling  coefficients  and  activity  thresholds
of  artificial  neuron  units
are  central  ingredients  in  neural  network  machines  simulating  the
brain  functions. Ricciardi  and  Umezawa[1] have
observed  that  in  the  study of  natural  brain it
is  pure  optimism  to  hope  to  determine  the  values of the  coupling
coefficients  and  the  activity  thresholds  of  all
neurons  by  means  of  anatomical  or  physiological  methods.
On the other hand, specific activities of the natural brain
persist  in  spite  of  changes  in
the  number  of  alive  neurons. In other words,
the  functioning  of  the  whole  brain appears  not  significantly
affected  by  the  functioning  of  the  single  neuron,  neither
physiological
observations  show  the  existence  of  special  long-lived  neurons  or
the  existence  of  a  large  redundancy  in  specialized  neuronal
circuits.  Besides  the  neurons,  many  other  thousands  of  elements, as
 glia  cells,
play a r\^ole  in  the  brain  activity, which, again, is not critically
dependent on the single cell functioning.

A characterizing
feature of the brain activity is instead related with nonlocality, namely
with the
existence  of  simultaneous  responces  in  several  regions
of  the  brain  to  some  external  stimuli.
This suggests  that the  brain
may  be  in  states  characterized  by  the  existence
of    long  range  correlations among its elementary constituents;
such   long  range  correlations seem to play a more fundamental r\^ole
than the functioning of the single cell in the brain activity.

Storing
and  recalling  information appear  as  a  diffuse
 activity  of  the  brain
not  lost  even  after  destructive  action  of  local
parts  of  the  brain  or  after  treatments  with  electric  shock  or
with  drugs.  This suggests to  model  these  memory  activities  as
{\it  coding}  of  the  brain  states whose  stability is  to be
derived  as  a  dynamical  feature rather than  as  a  property  of
specific  neural  nets  which  would  be  critically  damaged  by  the
above  destructive  actions.

Stable long  range  correlations  and
diffuse, nonlocal properties related  with  a  code specifying the system
state are dynamical features of {\it quantum} origin.
Ricciardi and Umezawa[1] have thus proposed
a quantum model where the  elementary  constituents  of
the  brain exhibit coherent
behaviour and macroscopic observables
are derived  as  dynamical  output  from their interaction.

Pioneering proposals relating advanced results in quantum optics, such as
holography, with brain
models were put forward by Pribram[3]. In more recent years, an analysis of
non-algorithmic and non-computational character of brain functions has been
made by Penrose[4], who has also proposed the quantum framework as the
proper one to bridge microscopic dynamics with macroscopic functional
activity of the brain.

For a general account of application of modern statistical mechanics
and spin glass theory to brain system see refs. [5]  and [6].

In the quantum model of Ricciardi and Umezawa
the  elementary  constituents are  not the
neurons  and  the  other  cells  and  physiological  units,  which  cannot
be  considered  as  quantum  objects,  but  some
dynamical  variables, called  corticons,  able  to
describe  stationary  or  quasi-stationary  states  of  the
brain.

A  crucial assumption, based on the fact that the brain is  an open
system  in  interaction
with  the  external  world, is that
information printing is achieved under the action of external  stimuli
producing breakdown of the continuous symmetry associated
to  corticons.

As well known, in spontaneously broken symmetry theories
the  Lagrangian is  invariant
under some group,  say $G$,  of  continuous  transformations; however,
the
minimum  energy  state,  i.e.  the  ground  state  or  vacuum,  of
the  system  is not invariant under the full group $G$, but under one of
its subgroups.
In  this  case,
general  theorems  of  QFT[7]  show  that  the  vacuum  is  an  ordered
state  and  collective  modes (called Nambu-Goldstone bosons) propagating
 over the  whole  system
are  dynamically  generated  and  are  the
carriers  of  the  ordering  information (long range correlations).  In
 other  words,  order
manifests  itself  as  a  global  property  dynamically
generated  and  the  quantum  numbers
characteristic  of  the  collective  mode  acts  as  {\it  coding}  for
the ground  state:  ordering and coding are  thus
achieved  by  the condensation of collective  modes
in  the vacuum.

One  important  point,  is  that  the
collective  mode
is  a  gapless  mode  and  therefore  its  condensation  in the vacuum
 does
not  add  energy to it.
 As a consequence,
the stability of the ordering and of the coding is insured. Another
consequence is that infinitely many vacua with different degrees of order
may exist, corresponding to different densities of the condensate.
In the infinite volume limit these vacua are each other
unitarily inequivalent and thus
represent possible
physical phases
of the system, which thus appears as a complex system with many macroscopic
configurations (phases).
The actual phase is determined once one among the many  degenerate
vacua is selected as an effect of some external action.

Transitions  among  these  vacua  are  in  general  not implementable
(non-existence of unitary transformations relating different vacua)
in  the
infinite  volume  limit;  however,  in the case of open  systems
these  transitions may  occur (phase transitions),
for  large  but  finite  volume,  due
to  coupling  with  external  environment.
The inclusion  of  dissipation
leads  thus  to  a  picture  of the system "living over many
ground  states" (continuously undergoing phase transitions)[8].
 It is interesting to observe that even very weak
(although above a certain threshold) perturbations may drive the system
through its macroscopic configurations[8].
In this way, occasional (random)
weak perturbations are recognized to play an important r\^ole in the
complex behavior of living systems.

The  observable  specifying  the  ordered state is
called  order  parameter and  acts  as  a  macroscopic variable
since  the  collective  modes  present
coherent dynamical  behavior. The order parameter is specific of the kind
of symmetry into play and may thus be considered as a code
specifying the vacuum.
The value of the order parameter is related with the density
of condensed Goldstone bosons in the vacuum and specifies the phase
of the system with relation to the considered symmetry. Since
physical properties are different for different phases, also the
value of
the order parameter may be considered as a code number
specifying the system state.
In conclusion, code numbers
specifying the phases may be organized in classes corresponding
to different kinds of dynamical symmetry.

A typical example of spontaneous breakdown of symmetry is provided by
the  ferromagnet  where  the  Lagrangian  is  invariant
under  the  spin  rotation group,  but  the  ground
state  is  invariant  only  under rotations
around  the
direction  of  the  magnetization.  The  collective  modes  are  the
spin-wave quanta or magnons  and  the  system phases are
indeed
macroscopically  characterized ({\it coded})  by  the  value  of  the
magnetization,
which  is  the  order  parameter.  The  magnetic  order  is  thus  a
diffused,  i.e.  macroscopic,  feature  of  the  system.

 The collective mode of the Ricciardi-Umezawa brain model has been called
symmetron[9] and the  information  storage  function  is
represented  by  the  coding  of  the  ground  state through
symmetron  condensation.

The  corticon  has  been
assumed[9] to  be  a  two  state
system  and  the  associated  symmetry is  a  phase
symmetry.

By following Fr\"ohlich[10], Del Giudice et al.[11-18]
have assumed that the
symmetry to be spontaneously broken in living matter is the rotational
symmetry for electrical dipoles.
Such an assumption is phenomenologically based on the fact that living
matter is made up by water
and other biomolecules equipped with electric dipoles. The
(electric) polarization
density thus plays the r\^ole of order parameter and the associated
Goldstone modes have been named dipole wave quanta ({\it dwq}).
In the QFT approach to living matter the dynamical generation of
collective modes thus shed some light on the problem of {\it change of
scale } in biological systems, namely the problem of the transition from
the microscopic scenario to the many macroscopic functional properties
(many macroscopic configurations) of
the living systems.
 Del
Giudice  et
al. have shown the superradiant  or  "lasering"
behaviour of  water  electrical  dipoles[15] and the self-focusing
propagation of the  electrical
field  in  ordered  water[13], thus providing a conjecture for the
formation of microtubules[13,18]. It has been shown[15] that the coherent
interaction of water molecules with the quantized radiation field leads
to a time scale for the coherent long range interaction much shorter
($10^{-14} sec $) than the one of short range interactions. Water
coherent domains are therefore protected from thermalization.
Solitary wave propagation on biomolecular chains, as proposed by
Davydov[19], has also been
studied[12] and related with triggering of  breakdown of symmetry.

 Spontaneous breakdown of electric dipole rotational symmetry has revealed
to be
useful also in
further developments of the quantum brain model (referred to as quantum
brain dynamics ({\it QBD}))  worked out by Jibu and Yasue[20-24] who have
identified the Ricciardi-Umezawa symmetron modes with {\it dwq} and the
corticon with the electric dipole field. They have obtained a first
understanding of anesthesia and have elaborated a formalism for the
superradiant propagation of electromagnetic field in cytoskeleton
microtubules, also in relation to computational functions possibly
associated to them[25].

Summing up, in the quantum brain  model external stimuli aimed to
information printing trigger the spontaneous breakdown of symmetry. The
stability  of  the
memory  is  insured  by  the  fact  that  coding  occurs  in  the  lowest
energy  state and  the  memory  nonlocal character  is  guaranteed  by  the
coherence  of  the {\it dwq} (or symmetron) condensate.

The  recall  process  is  described  as  the  excitation  of {\it dwq}
modes  under external  stimuli  of  a  nature  similar  to  the
ones producing the memory printing process.  When  the
{\it dwq}  modes  are  excited  the  brain  "consciously  feels"[9] the
pre-existing  ordered  pattern  in  the  ground  state.

Short-term  memory  is  finally  associated  to  metastable  excited
states  of  {\it dwq}  condensate[1].  For  a  discussion  on  this  point
see  also  ref.[26].

The  electrochemical  activity  observed  by
neurophysiology  provides,  according  to
Stuart et al.[9], a  first  responce  to  external
stimuli  which, through
some intermediate interaction, has  to  be
coupled  with  the dipole field (or corticon)  dynamics  so  to  allow
the  coding
of  the  ground  state. One possibility, according to QFT approach
to living matter[11-13], is that electrochemical activity may trigger, e.g.
through ATP reaction, solitary dipolar waves on biomolecular chains.
These solitary waves may in turn produce domains of nonzero polarization in
the surrounding
water molecules and the associated {\it dwq} condensation.
In the original brain model it  is
conjectured  that  the  formation  of ordered local domains  may  play  a
relevant  r\^ole  in  this intermediate coupling[9].

In the quantum brain model only one kind of symmetry is assumed (the dipole
rotational symmetry). Thus there is only one class of code numbers.
Suppose a
vacuum of specific code number has been selected by
the printing of a specific information. The brain then sets in that state
and no other vacuum state is successively accessible for recording another
information, unless a phase transition to the vacuum
specified by the new code number is
produced under the external stimulus carrying the new information. This
will destroy the previously stored information
({\it overprinting}): Vacua labelled by different code numbers are
accessible only through a sequence of phase transitions from one to
another one of them.

Such a problem of {\it memory capacity} was already mentioned by
Stuart et al.[9], who realized that the model was too simple
to allow the recording of a huge number of informations. Stuart et al. then
proposed that the model could be extended in such a way to present a huge
number of symmetries (a huge number of code classes)
and "a realistic model
would therefore require a vector space of extremely high
dimensions"[9], which however would introduce serious difficulties and
spoil
its practical use.

The purpose of the present paper is to show
that, by taking into account the fact that the brain is an open system with
dissipative dynamics, one may reach a
solution to the problem of memory capacity which does
not require the introduction of a huge number of symmetries.

It will be shown  that, even  by  limiting
the  analysis  to  one  kind  of  symmetry, infinitely many
vacua
are accessible to memory printing in a way
that
in a sequential information recording the successive printing of
information
does not destroy the previous ones, thus allowing a huge memory capacity.
Taking into account
dissipation is crucial in reaching such a result.

Section 2 is devoted to the presentation of the dissipative quantum brain
dynamics ({\it DQBD}). Its connection with thermal field theory and
squeezed coherent states is discussed in sections 3 and 4, respectively.

\bigskip

\medskip
\line{{\bf 2. Dissipative quantum dynamics of the brain}\hfill}

\medskip
\indent In this section the quantum model of the brain proposed by
Ricciardi and Umezawa is extended
to dissipative
dynamics by resorting to some results on
dissipative systems in QFT[2,27,28]. It will be shown that the problem
of memory capacity may have a solution in the framework of dissipative
quantum brain dynamics.

Let us start by the (trivial) observation that "only the past can be
recalled". This means that {\it memory printing breaks the time-reversal
symmetry of the brain dynamics} and is another way to express the (obvious)
fact that brain is an open, dissipative system coupled
with external world.

As a matter of fact, in the quantum brain model
spontaneous breakdown of dipole rotational symmetry is triggered by the
coupling of the
brain with external stimuli. Here, however, our attention is focused
on the fact that once the dipole rotational symmetry
has been broken
(and information has thus been recorded), then, {\it as a consequence},
time-reversal
symmetry is also broken: {\it Before} the information recording process,
the brain can in principle be in anyone of
the infinitely many (unitarily inequivalent) vacua.
{\it After} information has been recorded, the brain state is completely
determined
and the
brain cannot be brought to the state configuration in which it was {\it
before} the
information printing occurred  ({\it ...NOW you know it!...}).

Thus, information printing introduces
{\it the arrow of time} into brain dynamics.
Due to memory printing process time evolution of the brain states is
intrinsically irreversible.

Ricciardi and Umezawa[1] have studied
the brain non-stationary
or quasi-stationary states in the stationary approximation, thus
avoiding damped oscillations. In the following discussion,
on the contrary, we will consider non-stationary states without using the
stationary approximation.

A central feature of the quantum dissipation formalism[2,27,28]is the
duplication of the field describing the dissipative system.

Let
$a_{\kappa}$
and ${\tilde a}_{\kappa}$ denote the gapless {\it dwq}
mode and the doubled mode
required by canonical
quantization
of damped systems[2,27], respectively. $\kappa$ generically labels the
field degrees of freedom, {\it e.g.} spatial momentum. The
$\tilde a$ mode is the "time-reversed mirror image"[2]
of the $a$ mode and represents the environment mode.

The canonical commutation relations ({\it CCR})
of the bosonic $a_{\kappa}$ and ${\tilde a}_{\kappa}$ operators  are:
$$
[\, a_{\kappa} , a_{\lambda}^{\dagger}\, ] = \delta_{\kappa , \lambda} =
[\, {\tilde a}_{\kappa} , {\tilde a}_{\lambda}^{\dagger}\, ] \quad ; \quad
[\, a_{\kappa} , {\tilde a}_{\lambda}^{\dagger}\, ] = 0 = [\, a_{\kappa} ,
{\tilde a}_{\lambda}\, ] \quad . \eqno{(1)}
$$

It is convenient[2] to work with the bosonic operators $A_{\kappa}$ and
${\tilde A}_{\kappa}$. These are related to $a_{\kappa} $ and ${\tilde a}
_{\kappa}$ by the linear canonical transformations
 $ {A_{\kappa}  \equiv {1\over{\sqrt 2}}
( a_{\kappa}  +{\tilde a}_{\kappa}  )}$, $
{{\tilde A}_{\kappa}  \equiv {1\over{\sqrt 2}} ( a_{\kappa}  -
{\tilde a}_{\kappa} )}$, which indeed preserve the {\it CCR}'s:
$$
[\, A_{\kappa} , A_{\lambda}^{\dagger}\, ] = \delta_{\kappa , \lambda} =
[\, {\tilde A}_{\kappa} , {\tilde A}_{\lambda}^{\dagger}\, ] \quad ; \quad [\,
A_{\kappa} ,
{\tilde A}_{\lambda}^{\dagger}\, ] = 0 = [\, A_{\kappa} , {\tilde
A}_{\lambda}\, ] \quad .
\eqno{(2)}
$$
In the following, the ${\tilde A}$ modes will be called the "mirror
modes".

Let  $\{ | {\cal N}_{A} , {\cal N}_{\tilde A}  > \}$ be
the set of simultaneous
eigenvectors of ${\hat N}_{A} \equiv A^{\dagger} A$ and
${\hat N}_{\tilde A} \equiv {\tilde A}^{\dagger} {\tilde A}$,
with ${\cal N}_{A}$ and ${\cal N}_{\tilde A}$
non-negative integers and let $|0>_{0}  \equiv | {\cal N}_{A} = 0 , {\cal
N}_{\tilde A} = 0 >$
such that $A |0>_{0}  = 0 = B|0>_{0} $.

In ref.[2] it has been shown that the
quantum
dynamics of an (infinite) collection of
damped harmonic oscillators $A_{\kappa}$ is ruled by the Hamiltonian
$$
H = H_{0}   + H_{I} \quad , \eqno{(3a)}
$$
$$
 H_{0} = \sum_{\kappa} \hbar \Omega_{\kappa}  \bigl (
A_{\kappa}^{\dagger} A_{\kappa} - {\tilde A}_{\kappa}^{\dagger} {\tilde
A}_{\kappa} \bigr ) \quad ,
 \eqno{(3b)}
$$
$$
H_{I} = i \sum_{\kappa} \hbar \Gamma_{\kappa} \bigl (
A_{\kappa}^{\dagger} {\tilde A}_{\kappa}^{\dagger} - A_{\kappa} {\tilde
A}_{\kappa} \bigr ) \quad , \eqno{(3c)}
$$
where $\Omega_{\kappa}$ is the frequency and $\Gamma_{\kappa}$
 is the coupling constant.

 It is
interesting to observe that  in order to describe the dissipative system
one does not need the details of the environment: its
{\it effective} action on the system is globally represented by the action
of the "mirror image" of the system. We will comment
more on this point in the
following.

In order to take into account dissipativity we thus require that
the memory state is
a zero energy eigenstate of $H_{0}$ (vacuum) which therefore, at certain
initial time, say
$t_{0} = 0$, is a condensate of {\it equal number} of modes $A_{\kappa}$
and mirror modes ${\tilde A}_{\kappa}$ for any $\kappa$. Clearly, we then
have infinitely many memory states at $t_{0} = 0$, each one corresponding
to a different number ${\cal N}_{A_{\kappa}}$ of $A_{\kappa}$ modes, for
all ${\kappa}$, provided  $
 {\cal N}_{A_{\kappa}} - {\cal N}_{{\tilde A}_{\kappa}} = 0$ for all
 ${\kappa}$.

Let  ${|0> }_{\cal N}$ denote
the memory state  with ${\cal N}  \equiv
\{ {\cal N}_{A_{\kappa}}  =  {\cal N}_{{\tilde
A}_{\kappa}},  \forall  \kappa, at~~  t_{0} = 0 \}$
the
set of integers defining the "initial value" of the condensate,
namely the code number (or simply the code) associated to
the information recorded at time $t_{0} = 0$.

At finite volume $V$, the memory state $ {|0 >} _{\cal N} $
can be then represented
as a two-mode Glauber coherent state[29] (i.e. a generalized coherent state
for $su(1,1)$):
$$
{|0 >}_{\cal N}
= \exp{\bigl(-iG({\theta})\bigr)}|0>_{0}  = \prod_{\kappa}
{1\over{\cosh{{\theta}_{\kappa}}}} \exp{
\left (- \tanh {\theta_{\kappa} } J_{+}^{(\kappa )} \right )} |0>_{0}
 \quad , \eqno{(4)}
$$
with $ {J_{+}^{(\kappa
)} \equiv A_{\kappa}^{\dagger} {\tilde A}_{\kappa}^{\dagger}}$ and
$$
G({\theta}) =- i \sum_{\kappa} \theta_{\kappa} \bigl (
A_{\kappa}^{\dagger} {\tilde A}_{\kappa}^{\dagger} -A_{\kappa} {\tilde
 A}_{\kappa} \bigr ) \quad
 ~. \eqno{(5)}
$$

In Eq. (4) the $\cal N$-set,
${\cal N} \equiv
\{ {\cal N}_{A_{\kappa}} = {\cal N}_{{\tilde
A}_{\kappa}}, \forall \kappa, at~~ t_{0}=0 \}$,
is related to  $\theta
\equiv \{ {\theta}_{\kappa}\}$ by
$$
{\cal N}_{A_{\kappa}} =
 {_{\cal N}}< 0| A_{\kappa}^{\dagger} A_{\kappa}{|0>}
 _{\cal N} = \sinh^{2} \theta_{\kappa} ~, \eqno{(6)}
$$
and we will use the notation ${\cal N}_{A_{\kappa}}(\theta) \equiv
{\cal N}_{A_{\kappa}}$.

The $\theta$-set is
conditioned by the requirement that $A$
and $\tilde A$ modes
satisfy the
Bose distribution at time $t_{0} = 0$:
$$
{\cal N}_{A_{\kappa}}(\theta) = \sinh^{2}  \theta_{\kappa}
 = {1\over{{\rm e}^{\beta
 E_{\kappa}} - 1}} \quad , \eqno{(7)}
$$
where  $ {\beta {\equiv} {1\over{k_{B} T}}}$  denotes the inverse
temperature
at time $t_{0}=0$ ($k_{B}$ is the Boltzmann
constant).  We thus
recognize  $\{{|0>}_{\cal N}\}$  as a representation of
the {\it CCR}'s at finite temperature, equivalent
with the Thermofield Dynamics representation $\{ |0({\theta}(\beta) )>
\}$[30,31].

We note that ${|0>}_{\cal N}$
is normalized to $1$ for all $\cal N$
and that in the infinite volume limit  $\{{|0>}_{\cal N} \}$   and
$\{{|0 >}_{\cal N'}\}$ are representations of the {\it CCR}'s
each other unitarily inequivalent for different codes ${\cal N}  \neq {\cal
N'}$.
We have thus at $t_{0} = 0$ the splitting, or
{\it foliation}, of the space of states into infinitely many unitarily
inequivalent representations of the {\it CCR}'s.
{\it The freedom thus introduced by the degeneracy among the vacua
${|0>}_{\cal N}$, for all
$\cal N$, plays a crucial r\^ole in solving
the problem of memory capacity}.
A huge number of sequentially recorded informations may {\it coexist}
without destructive interference since infinitely many vacua
${|0>}_{\cal N} $  are independently accessible. Recording information of
code $\cal N'$
does not necessarily produce destruction of previously printed information
of code
${\cal N} \neq {\cal N'}$, contrarily to the nondissipative case,
where differently coded vacua are
accessible only through a sequence of phase transitions from one to
another one of them. In the present dissipative case the "brain
(ground) state" may be represented as the collection (or the
superposition) of the full set of memory states
${|0>}_{\cal N}$, for all
$\cal N$. Alternatively, one
may also think of the brain as a complex system with a
huge number of macroscopic states (the memory states).

In order to better clarify this point it is useful
to consider the dynamical group structure associated
with our system. For each ${\kappa}$,
the underlying group is
$SU(1,1)$[2]. Let us neglect
for the moment the suffix $\kappa$ for simplicity.
The two-mode
realization of the algebra $su(1,1)$ is generated by
$$
J_{+} = A^{\dagger}  {\tilde A}^{\dagger} \quad ,
\quad J_{-} = J_{+}^{\dagger} = A  {\tilde A}
\quad , \quad
J_{3} = {1\over{2}} (A^{\dagger} A +
 {\tilde A}^{\dagger}  {\tilde A} + 1)\quad ,
\eqno{(8)}
$$
$$
[\, J_{+} , J_{-}\, ] = - 2 J_{3} \quad ,
\quad [\, J_{3}  , J_{\pm}\, ] = \pm
J_{\pm} \quad . \eqno{(9)}
$$
\noindent The Casimir operator ${\cal C}$ is given by
$ {{\cal C}^{2} \equiv {1\over {4}} + J_{3}^{2} - {1\over{2}} (
J_{+} J{-}  + J_{-} J_{+} )}  =  {1\over{4}} ( A^{\dagger} A  -  {\tilde
 A}^{\dagger}  {\tilde A})^{2} $.
We thus see that the eigenstates of
$H_{0}$ can be expressed in terms of the basis of simultaneous
eigenstates of $\cal C$ and of $ {\left ( J_{3} -
{1\over{2}} \right )}$ in the representation labelled by the value $j \in
\ZzZ_{1\over{2}}$ of ${\cal C}$,  $\{ | j , m > \, ; \, m \geq |j| \}$ :
$$
\eqalign{
{\cal C} | j , m > &= j | j , m > \quad , \quad j = {1\over{2}}
({\cal N}_{A} - {\cal N}_{\tilde A}) \quad ; \cr \left ( J_{3} -
{1\over{2}} \right ) | j , m > &= m | j , m > \quad , \quad m =
{1\over{2}} ({\cal N}_{A} + {\cal N}_{\tilde A}) \quad . \cr} \eqno{(10)}
$$

The memory state corresponds to the choice $j = 0$ (for all $\kappa$) and
we see that,
at certain time $t$
there are (for each $\kappa$) $m$ {\it coexisting, independent} eigenstates
of $\cal C$ (of course, by reintroducing the $\kappa$ suffix,
we have
$m_{\kappa} = {\cal N}_{A_{\kappa}} = {\cal N}_{{\tilde A}_{\kappa}}$ and
the $m$-set, $m \equiv \{ m_{\kappa} \}$, corresponds to the
${\cal N}$-set).

As a result, the $SU(1,1)$ structure of the
dissipative dynamics introduces $m$-coded "replicas" of the system
(foliation of the state space) and
information printing can be
performed in each replica without destructive interference with recorded
informations in the other replicas. In the nondissipative case the "$m$-
freedom" is missing and consecutive
information printing produces {\it
overprinting}.

The non-existence in the infinite
volume limit of unitary transformation which may map one representation
of code ${\cal N}$ to another one of code ${\cal N'}$ guaranties that
the corresponding printed informations are indeed
{\it different} or {\it distinguishable} informations ( $\cal N$ is a
{\it good} code) and that each information printing is also protected
against interference from other information printing (absence of {\it
confusion} among
informations). The effect of finite (realistic) size of the system may
however spoil
unitary inequivalence and may lead to "association" of memories. We will
comment more on this point in section 4.

We have $[\, H_{0} , H_{I}\, ] = 0$.
 The commutativity of $H_{0}$  with ${H}_{I}$ ensures that
the number $
( {\cal N}_{A_{\kappa}} - {\cal N}_{{\tilde A}_{\kappa}})$
is a constant of motion
for any $\kappa$.

We therefore realize that in the dissipative dynamics
ruled by the Hamiltonian (3a),
although ${\cal N}_{A_{\kappa}}$ and ${\cal N}_{{\tilde A}_{\kappa}}$ are
allowed to separately change in time,
their difference is
kept constantly zero during time evolution.

Formally,
at finite
volume $V$,
the time evolution of the memory
state
${|0>}_{\cal N}$ is given by
$$
\eqalign{&
| 0(t) >_{\cal N} = \exp{\left ( - i t {{H}\over{\hbar}}\right )}
|0>_{\cal
N} = \exp{\left ( - i t {{H_{I} }\over{\hbar}}\right )}
 |0>_{\cal N} \cr &= \prod_{\kappa} {1\over{\cosh{(\Gamma_{\kappa} t
  - {\theta}_{\kappa} )}}} \exp{
\left ( \tanh {(\Gamma_{\kappa} t - {\theta}_{\kappa}  )} J_{+}^{(\kappa
)} \right )} |0>_{0}   \quad , \cr } \eqno{(11)}
$$
which is again  a generalized coherent state for
$su(1,1)$. In obtaining
Eq.(11) we used the commutativity between $H_{I}$ and
$G(\theta)$.

Let us observe that the vacuum $|0(t)>_{\cal N} $  is specified  by the
initial value ${\cal N}$,
 at $t_{0} = 0$, of  the condensate.

We note that $
 {}_{\cal N}<0(t) | 0(t)>_{\cal N}  =  1, \quad \forall t $ ,
and that, provided $ {\sum_{\kappa} \Gamma_{\kappa}
 > 0}$,
$$
\lim_{t\to \infty} {}_{\cal N}<0(t) | 0>_{\cal N}  \, \propto \lim_{t\to
\infty}
 \exp{\left ( -t  \sum_{\kappa}  \Gamma_{\kappa}  \right )} = 0 \quad .
 \eqno{(12)}
$$
Using the customary continuous limit relation $ {
\sum_{\kappa} \mapsto {V\over{(2 \pi)^{3}}} \int \! d^{3}{\kappa}}$, in the
infinite-volume limit we have (for $ {\int \! d^{3} \kappa \,
\Gamma_{\kappa}}$ finite and positive)
$$
\eqalign{
&{}_{\cal N}<0(t) | 0>_{\cal N}  \mapbelow{V \rightarrow \infty} 0 \quad
\forall \,
 t \quad , \cr
&{}_{\cal N}<0(t) | 0(t') >_{\cal N}  \mapbelow{V \rightarrow  \infty} 0
\quad \forall \, t\, , t' \quad , \quad t \neq t' \quad . \cr} \eqno{(13)}
$$

In the infinite volume limit,  time evolution of  $|0>_{\cal N}$
would be rigorously frozen according to Eq. (13) (
the states $|0(t)>_{\cal N} $ and the associated Hilbert spaces are each other
unitarily inequivalent for different time values $t \neq t'$ in the
infinite volume limit);
however, in
realistic situations, a finite life-time may be
possible due to effects of
the system boundaries (cf. Eq.(12)).

Time evolution of the memory state $|0>_{\cal N}$ is thus
represented as the trajectory of "initial condition" specified by the $\cal
N$-set in the space of the representations $\{ |0(t)>_{\cal N}  \}$ of
the {\it CCR}'s.
The non-unitary character of time-evolution implied by damping
is consistently recovered in the unitary inequivalence among
representations at different times in the infinite-volume limit.

We also have
$$
{}_{\cal N}<0(t) | 0>_{0} = \exp{\left ( - \sum_{\kappa}\ln
\cosh{(\Gamma_{\kappa} t
  -  {\theta}_{\kappa})}\right )}
   \quad ,
\eqno{(14)}
$$
which shows that at time $t = \tau$,
with $\tau$ the largest of the values ${\tau_{\kappa}}  \equiv {
{{\theta}_{\kappa}({\cal N}_{\kappa})}\over{\Gamma_{\kappa}}}$, the memory
state $|0>_{\cal N}$ is reduced (decayed)
to the "empty" vacuum $|0>_{0}$: the information has been {\it forgotten}.
At the time $t = \tau$ the state $|0>_{0}$ is available for recording
a new
information.

It is interesting to observe that in order to not completely forget
certain information,
one needs
to "restore" the ${\cal N}$ code,
which
corresponds to "refresh" the memory by {\it brushing up} the subject
(external stimuli maintained memory).

We observe that
the number of modes of type $A_{\kappa}$ is given, at each instant $t$ , by
$$
{\cal N}_{A_{\kappa}}(\theta,t) \equiv {}_{\cal N}< 0(t) |
A_{\kappa}^{\dagger} A_{\kappa} | 0(t) >_{\cal N}  =  \sinh^{2}\bigl (
\Gamma_{\kappa} t - {\theta}_{\kappa}   \bigr )
 \quad  \eqno{(15)}
$$
and similarly for the modes of type ${\tilde A}_{\kappa}$. Eq. (15)
shows
that the assigned initial condition is satisfied (cf. Eq.(6))and that, as
already observed above, the information code is washed out after a time
$t = \tau$.
We will
comment more on this point in the following section.

Finally, we note that at each $t$
$$
\eqalign{
&{1\over{\cosh{(\Gamma_{\kappa} t -
{\theta}_{\kappa}  )}}} A_{\kappa}^{\dagger} |0(t)>_{\cal N} \, =
{1\over{\sinh{(\Gamma_{\kappa} t -
{\theta}_{\kappa}  )}}} {\tilde A}_{\kappa} |0(t)>_{\cal N}  \quad ,
\cr  &{1\over{\cosh{(\Gamma_{\kappa} t - {\theta}_{\kappa}
 )}}} {\tilde A}_{\kappa}^{\dagger} |0(t)>_{\cal N} \, =
{1\over{\sinh{(\Gamma_{\kappa} t - {\theta}_{\kappa} )}}} A_{\kappa}
|0(t)>_{\cal N}  \quad , \cr} \eqno{(16)}
$$
which show that the creation of a mode $A_{\kappa}$ is equivalent to the
destruction of a mode ${\tilde A}_{\kappa}$ and vice-versa.  This leads us to
interpreting the ${\tilde A}_{\kappa}$ modes as the holes for the modes
$A_{\kappa}$ [2] (see also [30]).

In the following sections
we discuss the connection of the $DQBD$
with thermal field theory and squeezed coherent states.

\bigskip
\medskip
\line{{\bf 3. Thermal field theory}\hfill}

\medskip

The state $|0(t)>_{\cal N} $  may be written as[2,30]:
$$
|0(t)>_{\cal N} \, = \exp{\left ( - {1\over{2}} S_{A} \right )}
|\,{\cal I}>\, = \exp{\left ( - {1\over{2}} S_{\tilde A} \right )}
|\,{\cal I}> \quad ,  \eqno{(17)}
$$
where
$
|\,{\cal I}>\, \equiv \exp {\left( \sum_{\kappa}
A_{\kappa}^{\dagger}
{\tilde A}_{\kappa}^{\dagger} \right)} |0>_{0}$ and
$$
S_{A} \equiv - \sum_{\kappa} \Bigl \{ A_{\kappa}^{\dagger} A_{\kappa}
\ln \sinh^{2} \bigl ( \Gamma_{\kappa} t - {\theta}_{\kappa}  \bigr ) -
A_{\kappa} A_{\kappa}^{\dagger} \ln \cosh^{2} \bigl ( \Gamma_{\kappa} t -
{\theta}_{\kappa}\bigr ) \Bigr \} \quad . \eqno{(18)}
$$
$S_{\tilde A}$ is given by an
expression similar to (18) with ${\tilde A}_{\kappa}$ and ${\tilde
A}_{\kappa}^{\dagger}$ replacing $A_{\kappa}$ and
$A_{\kappa}^{\dagger}$,
respectively.  Since $A_{\kappa}$ and ${\tilde A}_{\kappa}$ commute
(see (2)), we shall simply write $S$ for either $S_{A}$ or
$S_{\tilde A}$. It is known[2,30] that $S$ can be interpreted as the
entropy operator for the dissipative system.

Note that $|0(t)>_{\cal N} $ depends on time
only through the exponential of  $ { {1\over{2}} S_{A}}$
(or respectively,  $ { {1\over{2}} S_{\tilde A}}$) whose
operatorial part depends uniquely on the $A$ ($\tilde A$) variables: thus
Eq. (17) may be regarded as the projection on the (sub)system $A$ ($\tilde
A$)
with the elimination of the $\tilde A$ ($A$) variables.

The time variation of $|0(t)>_{\cal N} $ at finite volume $V$ is given by$^{\,
[2,28]}$,
$$
{{\partial}\over{\partial t}} |0(t)>_{\cal N}  =  - \left ( {1\over{2}}
{{\partial S}\over{\partial t}} \right ) |0(t)>_{\cal N}  \quad . \eqno{(19)}
$$
which shows that $ {i \left ( {1\over{2}} \hbar {{\partial
S}\over{\partial t}} \right )}$
is the generator of time-translations,
namely time-evolution is controlled by the entropy variations. It is
an interesting feature of the present treatment of dissipation that
the same operator $S$ that
controls time evolution also defines the dynamical variable
whose expectation value is formally the entropy:
this feature indeed reflects the
irreversibility of time evolution (breakdown of time-reversal symmetry)
characteristic of dissipative systems, namely
the choice of a privileged
direction in time evolution ({\it arrow of time}).

In order to study the stability condition to be satisfied at each time $t$
by the state $|0(t)>_{\cal N} $ let us introduce the  free energy
functional[2,30]
$$
{\cal F}_{A} \equiv {}_{\cal N}<0(t)| \Bigl ( H_{A} -{1\over{\beta}}
S_{A} \Bigr ) |0(t)>_{\cal N}  \quad . \eqno{(20)}
$$
$\beta$ is a strictly positive function of time representing
the inverse temperature: $ {\beta
(t) = {1\over{k_{B} T(t)}}}$; $H_{A}$ is the part of $H_{0}$ relative to
the $A$-modes only, namely $ {H_{A} = \sum_{\kappa} \hbar \Omega_{\kappa}
A_{\kappa}^{\dagger} A_{\kappa}}$.
Let $\Theta_{\kappa} \equiv \Gamma_{\kappa} t - \theta_{\kappa}$ and
$E_{\kappa} \equiv \hbar \Omega_{\kappa}$.
The
stationarity condition
$$
{{\partial {\cal F}_{A}}\over{\partial \Theta_{\kappa}}} = 0 \quad , \quad
\forall \kappa \quad ,  \eqno{(21)}
$$
gives $
\beta (t) E_{\kappa} = - \ln \tanh^{2} (\Theta_{\kappa})$. This
finally leads to
$$
{\cal N}_{A_{\kappa}}(\theta,t) = \sinh^{2} \bigl
( \Gamma_{\kappa} t - {\theta}_{\kappa}\bigr ) =
{1\over{{\rm e}^{\beta (t) E_{\kappa}} - 1}} \quad , \eqno{(22)}
$$
\noindent which is the Bose distribution for
$A_{\kappa}$ at time $t$.

 Again, this allows us to recognize $\{ |0(t)>_{\cal N}  \}$ as a
representation of
the {\it CCR}'s at finite temperature, equivalent
with the Thermofield Dynamics representation[30,31].

One can see (cf. Eq. (18)) that the the entropy ${\cal S}(t) =
<0(t)| S |0(t)>_{\cal N}  $
is a
decreasing function of time in the interval
$(t_{0} = 0, \tau )$ meaning that the memory state, although not conserved
in time,
is however "protected" from "going back" to the "unrecorded" or "blank"
vacuum state
(memory cancellation).
Of course, here it is crucial the energy exchange with the environment and
we are also assuming finite volume effects. In the infinite
volume limit, as already noticed, time evolution would be frozen and
stability rigorously ensured.
  One can also see that the entropy,
for both $A$ and $\tilde A$ system,
grows monotonically with $t$ from value $0$ at  $t = \tau$ to
infinity at $t = \infty$ . However, the
difference ~$(S_{A} - S_{\tilde A})$~
is constant in time: $
[\, S_{A} - S_{\tilde A} , H\, ] = 0 \quad .$
Since the $\tilde A$-particles are the holes for
the $A$-particles, ~$(S_{A} -S_{\tilde A})$~ is, in fact, the (conserved)
entropy for the
complete system.

Also, it can be shown[2,31] that, as time evolves, the change in the energy
$ {E_{A} \equiv \sum_{\kappa} E_{\kappa} {\cal N}_{A_{\kappa}}}$  and in
the entropy is given by
$$
d E_{A} = \sum_{\kappa} E_{\kappa} \dot{\cal N}_{A_{\kappa}} d t =
 {1\over{\beta}} d {\cal S}_{A}  \quad ,
  \eqno{(23)}
$$
i.e.
$$
d E_{A} - {1\over{\beta}} d {\cal S}_{A} = 0 \quad . \eqno{(24)}
$$
When $ {{{\partial \beta}\over{\partial t}} = - {1\over{k_{\tilde A}
T^{2}}} {{\partial T}\over{\partial t}} \approx ~0}$, namely
changes in inverse temperature are slow,
eq. (24) can directly be obtained by minimizing the free energy (20): $
d {\cal F}_{A} = d E_{A} - {1\over{\beta}} d {\cal S}_{A} = 0 $.
$E_{A}$ is thus recognized as the internal energy of the system. Eq. (24)
also expresses the
first principle of thermodynamics for a system coupled with environment
at constant temperature and in absence of mechanical work.
One may define as usual heat as $ {dQ={1\over{\beta}} dS}$. Thus the
change in time of condensate (Eq. (23))
turns out into
heat dissipation $dQ$.

In conclusion, time evolution of the  $ {\cal N}$-coded
memory state is represented as a trajectory
of initial condition  ${\cal N} = \{{\cal
N}_{A_{\kappa}}\}$ running over the space of the representations
$\{ |0(t)>_{\cal N}  \}$,
each one minimizing the free energy functional.

We close this section by observing that
the time evolution above discussed is different from the quantum decay
process of memory involving the virtual dynamics of instantons
which is discussed
in ref. [23].
Dissipation has been treated above as a realistic physical feature of
the brain considered as an open system.
 The memory states thus
obey a truly dissipative dynamics not considered in [23].
The virtual dynamics
of instantons with quantum fluctuations due to tunnel effect can also be
considered in the present  dissipative framework
in the same fashion as it has been studied by Jibu ad Yasue in ref. [23].
However, it
deals with QFT features which are different from the ones presented in this
paper.

\bigskip
\medskip
\line{{\bf 4. Squeezing and concluding remarks}\hfill}

\medskip

We now briefly discuss the relation between
the memory states and the
squeezed coherent states which emerges in the dissipative dynamics above
presented.

It is easy to show that the operator
$\exp{\bigl(-iG({\theta})\bigr)}$
( see Eq.
(5))
is rewritten
in terms of the operators $a$ and ${\tilde a}$ as

$$
\exp {\bigl( -iG(\theta) \bigr) } =
\prod_{\kappa} {{ \exp{ \biggl( - {{\theta}_{\kappa} \over{2}}
\bigl( {a_{\kappa}}^{2} - {a_{\kappa}}^{\dagger 2} \bigr) \biggr) }}
{ \exp{  \biggl( {{\theta}_{\kappa} \over{2}}
\bigl( {{\tilde a}_{\kappa}}^{2} - {{\tilde a}_{\kappa}}^{\dagger 2}
\bigr) \biggr) } }}
$$


$$
{ \equiv} {{\prod}_{\kappa}}
{\hat S}_{a}( \theta_{\kappa}  ){\hat S}_{\tilde a}
(-\theta_{\kappa}  )  \quad ,
 \eqno{(25)}
$$
with $\hat {S_{a}}( \theta_{\kappa}  ) {\equiv} \exp\bigl(-{{
{\theta}_{\kappa}  }\over{2}}\bigl({a_{\kappa}}^{2} -{a_{\kappa}}^{
{\dagger} 2}\bigr)\bigr)$ and similar expression for ${\hat
{S_{\tilde a}}}(
-{\theta}_{\kappa})$ with ${\tilde a}$ and ${\tilde a}^{\dagger}$ replacing
$a$ and $a^{\dagger}$, respectively.

The operators ${\hat S}_{a}({ \theta}_{\kappa}  )$ and ${\hat
S}_{\tilde a}(
-{\theta}_{\kappa})$ are the squeezing operators
for the $a_{\kappa}$ and the ${\tilde a}_{\kappa}$ modes, respectively,
as well known in quantum optics[32]. The set $\theta {\equiv}
 \{ \theta_{\kappa} \}$ as well as each
 $ \theta_{\kappa} $ for all $\kappa$ is called the squeezing parameter.

An expression similar to Eq. (25), but with ${ \theta}_{\kappa}$ replaced
by $-{ \Gamma}_{\kappa} t$, is obtained for the time evolution operator
 $ {\cal U}
\equiv \exp{\left ( -i t
{{H_{I}} \over {\hbar}}\right )}$ (cf. Eq. (3c)).

{}From Eqs.(4) and (11) we thus conclude that the memory state and its time
evoluted state
are squeezed coherent states.

To illustrate the effect of the squeezing,
 let us focus our attention only on the
$a_{\kappa}$ modes for sake of definiteness.
For the $\tilde a$ modes we can
proceed in a similar way.

As usual, for given $\kappa$ we express the $a$
mode in terms of conjugate
variables of the corresponding oscillator.
By using dimensionless quantities
we thus write $a = X + iY$, with $[X,Y] = {i\over{2}}$. The uncertainty
relation is ${\Delta }X {\Delta}Y = {1\over{4}}$ , with
${{\Delta }X}^{2} = {{\Delta}Y}^{2} = {1\over{4}}$  for (minimum
uncertainty) coherent states. The squeezing occurs when
${{\Delta }X}^{2}
< {1\over{4}}$ {\it and} ${{\Delta}Y}^{2} > {1\over{4}}$ (or
${{\Delta }X}^{2}
> {1\over{4}}$ {\it and} ${{\Delta}Y}^{2} < {1\over{4}}$) in such a way
that the uncertainty relation remains unchanged.
Under the action of
$\exp{\bigl(-iG({\theta})\bigr)} $
the variances ${\Delta }X$  and ${\Delta}Y$
are indeed squeezed as
$$
{{\Delta }{X}}^{2}(\theta )
  = {{\Delta }
 {X}}^{2}\exp(2{\theta})~~,~~~~
  {{\Delta }{Y}}^{2}(\theta )
   =   {{\Delta }
  {Y}}^{2}\exp(-2{\theta} )~~.
  \eqno {(26a)}
$$

For the tilde-mode similar relations are obtained for the corresponding
variances:
$$
{{\Delta }{\tilde X}}^{2}(\theta )
  = {{\Delta }
 {\tilde X}}^{2}\exp(-2{\theta})~~,~~~~
  {{\Delta }{\tilde Y}}^{2}(\theta )
   =   {{\Delta }
  {\tilde Y}}^{2}\exp(2{\theta} )~~.
  \eqno {(26b)}
$$

For positive $\theta$, squeezing then
reduces the variances
of the $Y$ and $\tilde X$ variables, while the variances of the $X$
and $\tilde Y$ variables grow by the same amount so to keep the
uncertainty relations unchanged. This reflects, in terms of the $A$ and
$\tilde A$ modes, the constancy of the difference
${\cal N}_{A_{\kappa}} - {\cal N}_
{{\tilde A}_{\kappa}}$ against separate,
but equal, changes of ${\cal N}_{A_{\kappa}}$ and ${\cal
N}_{{\tilde A}_{\kappa}}$ (degeneracy of the memory states labelled by
different codes).

In conclusion, the memory code of a specific
information, namely the  $\theta$-set
$\{ \theta_{\kappa}({\cal N}_{\kappa}) \}$ (cf. Eq.(6)) , is nothing but
the squeezing parameter classifying the squeezed coherent states in the
hyperplane $(X,{\tilde X} ; Y, {\tilde Y})$. Note that to different
squeezed states (different
 $\theta$-sets) are associated unitarily inequivalent representations of
 the {\it CCR}'s  in the infinite volume limit: in dissipative quantum
brain dynamics the huge (infinite) number of squeezed states, labelled by
the
squeezing parameter  $\theta \equiv \{ \theta_{\kappa} \}$,
constitute the memory capacity.

As observed above, time evolution also contributes to squeezing. We find
$$
\eqalign{
 {{\Delta }X}^{2}({\theta}, t) & =  {1\over{4}} \exp(-2({\Gamma} t
 - {\theta}) ) \quad , \cr
  {{\Delta }Y}^{2}({\theta}, t)  & =  {1\over{4}}\exp(2({\Gamma} t
  - {\theta}) )~~. \cr}
 \eqno {(27)}
$$
Similar relations hold for the tilde-mode (with the exponential factors
exchanged).
In Eq. (27)
we used the values for the variances for the vacuum
$ |0>_{0} \equiv |0,
{\tilde 0}>$.

Eqs. (27) (and the corresponding ones for the tilde-mode) show the time
behaviour of the squeezing and we can recover the
analysis of time evolution made in the previous sections. Again, we note
the r\^ole of the characteristic time $\tau \equiv {
{\theta}\over{\Gamma}}$. We also see that
in the limit $t  \rightarrow  \infty$ the variances of the variables
$Y$ and ${\tilde X}$ become infinity making them completely spread out.

Let us now summarize the main points of the discussion presented in this
paper.

In dissipative quantum brain dynamics infinitely
many vacua coexist and a huge number of informations may be
sequentially recorded without destructive interference.

The problem of memory capacity in the
quantum brain model,
arising from the fact that vacua labelled by different
code numbers belonging to the same class are accessible only by a
sequence of phase transitions,
finds a solution in the intrinsic dissipative
character of brain dynamics.

As we have indeed stressed, the process
of information printing by itself produces the breakdown of time-reversal
symmetry and thus introduces the arrow of time into brain dynamics.
The key point is that
the resulting dissipative dynamics cannot be worked out without the
the introduction
of the time-reversed image (the tilde-system) of the original system.
As a consequence, energy degeneracy is introduced and the brain
ground state may be represented as a collection (or superposition)
of infinitely many degenerate vacua or memory states, each of them
labelled by a different code number and each of them independently
accessible to
information printing (without reciprocal interference).
Many information storage levels may then
coexist thus allowing a huge memory
capacity.

Differently stated, the brain system may be viewed as a complex system
with (infinitely) many macroscopic configurations (the memory states).
Dissipation, which is intrinsic to the brain dynamics, is recognized to be
the root of such a complexity, namely of the huge memory capacity.

Time evolution of the  $ {\cal N}$-
coded
memory state is represented as a trajectory
of initial condition  ${\cal N} = \{{\cal
N}_{A_{\kappa}}\}$ running over the states
$|0(t)>_{\cal N}$,
each one minimizing the free energy functional.

Memory states have also been shown to be squeezed coherent states.

Let us close the paper with few more comments.

The QFT approach to
living matter does not require the introduction of other symmetries than
the dipole rotational symmetry (and the electromagnetic
gauge symmetry, see
ref. [13]).
In the case other symmetries could be required in future developments of
such
an approach, to each broken symmetry will be associated a code class.
Then the memory state will carry as many labels (codes) as many dynamical
symmetries are broken. In such a case, the Goldstone modes associated to a
specific label may interfere with the Goldstone modes associated to some
other label of the same state. This may produce fluctuations in their
condensation and thus originates the mechanism of "association" of
memories, by which some information is recorded with some
"confusion" due to the presence of elements belonging to a different
information; or also, the recalling of some information  may trigger the
recalling of some other information.

As already observed,
association of memories may also occur when, as in the present paper, only
one kind of symmetry
is considered. In such a case,
"interferences" are due to the realistic (finite) size of the
system (boundaries effects) making the memory states
not exactly orthogonal (unitary  non-equivalence is spoiled).

We remark that the memory state is not invariant under
 $H_{I}$ (see Eq.(11)), while the Hamiltonian $H$ commutes with  $H_{I}$.
 Therefore, in
addition to breakdown of time-reversal (discrete) symmetry,
already mentioned
in the previous sections, we also have spontaneous breakdown of time
translation (continuous) symmetry.
Dissipation (i.e. energy
non-conservation) has been thus described in this paper
(see also ref. [2])
 as an effect of
breakdown of time translation and time-reversal symmetry.

Finally, according to the original
quantum brain model,
the recall process is described as the excitation of {\it dwq} modes
under an external stimulus which is
"essentially a replication signal"[9] of
the one responsible for
memory printing. When {\it dwq} are excited
the brain  "consciously feels"[9]
the presence of the
condensate pattern in the corresponding coded vacuum.
The replication signal thus acts as a probe by which the brain
"reads" the printed information.

In this connection we observe  that
the {\it dwq} may acquire an effective nonzero mass due to the effects of
the system finite size[12]. Such an effective mass will then
introduce a threshold in the excitation energy of {\it dwq} so that,
in order to trigger the recall process
an energy supply equal or greater than such a threshold is required.
Non sufficient energy supply may be experienced as a "difficulty in
recalling".
At the same time, however, the
threshold may positively act as a "protection" against unwanted
perturbations (including thermalization) and cooperate to the memory
state stability.
In the opposite case of zero threshold any replication signal could excite
the recalling  and the brain would fall in a state of "continuous flow of
memories".

As for the the "replication signal", it is interesting to observe that in
DQBD the $\tilde A$ system
 is indeed a "replication" of the $A$ system and plays in fact a central
 r\^ole in the recalling process: Eqs.(16) show that the creation
 (excitation)
of the $A$ mode is equivalent,
up to a factor, to the destruction (from the memory state) of the $\tilde
A$ mode. In this sense the coupling term of the $\tilde A$  mode with the
$A$ mode in the Hamiltonian can be seen as a self-interaction term
of the $A$ system, thus confirming the  r\^ole of
$\tilde A$ system in "self-recognition"
processes.

Remarkably, the tilde-system also
represent the environment effects and cannot be neglected since the brain
is an open system. Therefore the tilde-modes can never be eliminated from
the brain dynamics:
the tilde-modes thus might play a r\^ole as well
in the unconscious brain activity.
This may provide an answer to the question "as
whether symmetron modes would be required to account for unconscious brain
activity"[9].

Moreover, we have seen that
the $\tilde A$ system is the time-reversed
image
of the $A$ system. Thus the $\tilde A$ system is the "mirror in time"
system.
This fact, together with the r\^ole of the $\tilde A$ modes in the
self-recognition processes, leads us
to conjecture
(also accepting the literary image of
consciousness as a "mirror")
that tilde-system is actually responsible for consciousness mechanisms:
Consciousness emerges as a manifestation of the
dissipative quantum dynamics of the brain.

\medskip
\medskip

\line{{\bf Acknowledgments}\hfill}

I would like to warmly thank  H.Umezawa for many stimulating discussions
on the fascinating subject of brain dynamics.
I am also glad to thank E. Del Giudice, M. Jibu,
R. Penrose, K.H. Pribram, M. Rasetti and K.Yasue for the stimulating inputs
I received from them. Finally, I am grateful to M. Rasetti for his
encouragement in writing this paper.

\vfill\eject

\line{{\bf References}\hfill}
\medskip

\item{[1]} L.M. Ricciardi and H.Umezawa, {\it Kibernetik} {\bf 4}, 44
(1967) \smallskip
\item{[2]} E. Celeghini, M.
Rasetti and G. Vitiello, {\it Annals of Physics} (N.Y.) {\bf 215}, 156
(1992)
\smallskip
\item{[3]} K.H. Pribram, in {\it Macromolecules and behavior}, J.Gaito
ed., Academic Press,  N.Y.  1966.
\smallskip

\item{} K.H. Pribram, {\it Languages of the brain}, Englewood Cliffs, New
Jersey, 1971

\smallskip
\item{} K.H. Pribram, {\it Brain and perception}, Lawrence Erlbaum, New
Jersey, 1991

\smallskip

\item{[4]} R. Penrose  {\it The Emperor's new mind}, Oxford
University Press,  London 1989
\smallskip
\item{} R. Penrose,  {\it Shadows of the mind}, Oxford University
Press,  London 1993
\smallskip

\item{[5]} M. M\'ezard, G. Parisi and M. Virasoro, {\it Spin glass theory
and beyond}, World Sci.,  Singapore 1993

\smallskip
\item{[6]} D.J. Amit {\it Modeling brain functions},
Cambridge University Press,  Cambridge 1989

\smallskip
\item{[7]} C. Itzykson and J.B. Zuber, {\it Quantum Field Theory}, McGraw-
Hill, N.Y. 1980
\smallskip

\item{[8]} E. Del Giudice, R. Ma\'{n}ka, M. Milani and G. Vitiello,
{\it Phys.\ Lett.} {\bf 206B}, 661 (1988)
\smallskip
\item{} E. Celeghini, E. Graziano and G. Vitiello,
{\it Phys. Lett.} {\bf 145A}, 1 (1990)

\smallskip
\item{[9]} C.I.J. Stuart, Y. Takahashi and H. Umezawa, {\it J. \ Theor. \
Biol.} {\bf 71}, 605 (1978)

\smallskip
\item{} C.I.J. Stuart, Y. Takahashi and
H. Umezawa, {\it Found. Phys.} {\bf 9}, 301 (1979)

\smallskip

\item{[10]} H. Fr\"{o}hlich, {\it J.Quantum Chemistry} {\bf 2}, 641 (1968)
\smallskip
\item{} H. Fr\"{o}hlich, {\it Riv. Nuovo Cimento} {\bf 7}, 399 (1977)
\smallskip
\item{} H. Fr\"{o}hlich, {\it Adv. Electron. Phys.} {\bf 53}, 85 (1980)
\smallskip
\item{} H. Fr\"{o}hlich, in {\it Biological coherence and
response to external stimuli}, H. Fr\"{o}hlich ed., Springer-Verlag,
Berlin 1988, p.1
\smallskip
\item{[11]} E. Del
Giudice, S. Doglia, M. Milani and G. Vitiello, {\it Phys.\ Lett.} {\bf 95A},
508 (1983)
\smallskip
\item{[12]} E. Del Giudice, S. Doglia, M. Milani and G.
Vitiello, {\it Nucl.\ Phys.} {\bf B251 [FS 13]}, 375 (1985)

 \smallskip
\item{[13]} E. Del
Giudice, S. Doglia, M. Milani and G. Vitiello, {\it Nucl.\ Phys.} {\bf B275
[FS
17]}, 185 (1986)

\smallskip
\item{[14]} E. Del Giudice, S. Doglia, M. Milani and G.
Vitiello, in {\it Biological coherence and response to external stimuli},
H. Fr\"{o}hlich ed., Springer-Verlag, Berlin 1988, p.49
\smallskip
\item{[15]} E. Del
Giudice, G.
Preparata and G. Vitiello, {\it Phys.\ Rev.\ Lett.} {\bf 61}, 1085 (1988)
\smallskip
\item{[16]} G. Vitiello, {\it Nanobiology} {\bf 1}, 221 (1992)

\smallskip
\item{[17]} G. Vitiello, {\it Living matter physics}
 in {\it Coherent and
emergent phenomena in biomolecular systems}, P.A. Hanson, S. Hameroff, S.
Rasmussen and J.A. Tuszy\'nski eds., MIT-Press, in print.

\smallskip
\item{[18]} E. Del Giudice, S. Doglia, M. Milani and
G. Vitiello, in {\it Interfacial phenomena in biological systems}, M.
Bender ed., Marcel Dekker Inc., N.Y. 1991, p.279

\smallskip

\item{[19]} A.S.
Davydov, {\it Biology and quantum mechanics}, Pergamom, Oxford 1982
\smallskip
\item{}
A.S. Davydov, {\it Physica Scripta} {\bf 20}, 387 (1979)
\smallskip

\item{} A.S. Davydov and
N.I. Kislukha, {\it Sov. \ Phys.\ JEPT} {\bf 44},
571 (1976)

\smallskip
\item{[20]} M. Jibu  and K. Yasue,  in {\it Cybernetics and System
Research}, R. Trappl ed., World Scientific, London 1992, p.797
\smallskip
\item{[21]} M. Jibu  and K. Yasue, in {\it Proceed. of the First
Appalachian Conference on Behavioral Neurodynamics}, K.H.
Pribram ed., Radford University, Radford 1992
\smallskip
\item{[22]} M. Jibu  and K. Yasue,  {\it Cybernetics and Systems: An
International Journal}  {\bf 24}, 1 (1993)

\smallskip
\item{[23]} M. Jibu  and K. Yasue, in {\it Nature, Cognition and System
III}, E. Carvallo ed., Kluver Academic, London 1993
\smallskip
\item{[24]} M. Jibu , S. Hagan, S.R. Hameroff, K. H. Pribram and  K. Yasue,
{\it BioSystems} {\bf 32}, 195 (1994)
\smallskip

\item{[25]} S.R. Hameroff and R.C. Watt, {\it J. Theor. Biol.}
{\bf 98}, 549 (1982)

\smallskip

\item{} S.R. Hameroff, in
{\it Biological coherence and response to external  stimuli}, H.
Fr\"{o}hlich ed., Springer-Verlag, Berlin 1988, p.242

\smallskip
\item{[26]} S.Sivakami and V. Srinivasan, {\it J. Theor. Biol.} {\bf 102},
287 (1983)
\smallskip

\item{[27]} H. Feshbach and Y. Tikochinsky,  {\it Transact. N.Y. Acad. Sci.}
{\bf 38} (Ser. II), 44  (1977)
\smallskip

\item{[28]} S. De Filippo and G. Vitiello,  {\it Lett. Nuovo Cimento} {\bf
19},
 92 (1977)
\smallskip
\item{[29]}  J.R. Klauder and E.C. Sudarshan, {\it Fundamentals of Quantum
Optics}, Benjamin, New York, 1968
\smallskip

\item{[30]} Y. Takahashi and H. Umezawa, {\it Collective Phenomena} {\bf
 2}, 55 (1975)

\smallskip
\item{} H. Umezawa, H. Matsumoto and M. Tachiki, {\it Thermo
field dynamics and condensed states}, North-Holland, Amsterdam 1982

\smallskip
\item{[31]} H. Umezawa,
{\it Advanced field theory: micro, macro and thermal concepts}, American
Institute of Physics, N.Y. 1993

\smallskip
\item{[32]} H.P. Yuen, {\it Phys. Rev.} {\bf A13}, 2226 (1976)

\smallskip
\item{}  D. Stoler, {\it Phys. Rev.} {\bf D 1}, 3217 (1970)

\vfill\eject

\bye